\documentclass[letter]{aa} 

\usepackage{graphicx}
\usepackage{caption}
\usepackage{subcaption}
\usepackage{epstopdf}
\usepackage{amsmath}
\usepackage[english]{babel}
\usepackage{amssymb}
\usepackage{wrapfig}
\usepackage{pdflscape}
\usepackage{lscape}
\usepackage{longtable}
\usepackage{appendix}
\usepackage{textcomp}
\usepackage{multirow}
\bibpunct{(}{)}{;}{a}{}{,}
\usepackage{array}
\usepackage{ragged2e}
\usepackage{adjustbox}
\usepackage{txfonts}

\begin{document} 

   \title{SDSS J211852.96-073227.5: The first non-local, interacting, late-type intermediate Seyfert galaxy with relativistic jets\thanks{This paper includes data gathered with the 6.5-meter Magellan Telescopes located at Las Campanas Observatory, Chile.}}

\titlerunning{J2118-0732: an intermediate Seyfert with relativistic jets}

   \author{E. J\"{a}rvel\"{a}\inst{1}\thanks{\email{ejarvela@sciops.esa.int}}
           \and
          M. Berton\inst{2,3}
          \and 
          S. Ciroi\inst{4}
          \and
          E. Congiu\inst{5}
          \and
          A. L\"ahteenm\"aki\inst{3,6}
          \and
          F. Di Mille\inst{5}
          }

\authorrunning{E. J\"{a}rvel\"{a} et al.}

   \institute{European Space Agency, European Space Astronomy Centre, C/ Bajo el Castillo s/n, 28692 Villanueva de la Cañada, Madrid, Spain
        \and
            Finnish Centre for Astronomy with ESO (FINCA), University of Turku, Vesilinnantie 5, FIN-20014 University of Turku, Finland
        \and
            Aalto University Mets\"ahovi Radio Observatory, Mets\"ahovintie 114, FIN-02540 Kylm\"al\"a, Finland
        \and
            Dipartimento di Fisica e Astronomia ”G. Galilei”, Universit{\`a} di Padova, Vicolo dell’Osservatorio 3, 35122 Padova, Italy
        \and
           Las Campanas Observatory - Carnegie Institution for Science, Colina el Pino, Casilla 601, La Serena, Chile
        \and
            Aalto University Department of Electronics and Nanoengineering, P.O. Box 15500, FIN-00076, Aalto, Finland
             }

   \date{Received ; accepted }


  \abstract
   { 
   It has been often suggested that a tangible relation exists between relativistic jets in active galactic nuclei (AGN) and the morphology of their host galaxies. In particular, relativistic jets may commonly be related to merging events. Here we present for the first time a detailed spectroscopic and morphological analysis of a Seyfert galaxy, SDSS J211852.96-073227.5, at $z = 0.26$. This source has previously been classified as a gamma-ray emitting narrow-line Seyfert 1 galaxy. We re-observed it with the 6.5m Clay Telescope and these new, high-quality spectroscopic data have revealed that it is actually an intermediate-type Seyfert galaxy. Furthermore, the results of modelling the $Ks$-band near-infrared images obtained with the 6.5m Baade Telescope indicate that the AGN is hosted by a late-type galaxy in an interacting system, strengthening the suggested connection between galaxy interactions and relativistic jets. 
}
   

   \maketitle


\section{Introduction}
\label{sec:intro}

Relativistic jets, present in about 10\% of active galactic nuclei \citep[AGN; ][]{2017padovani2}, were long believed to be an exclusive property of giant elliptical galaxies that are host to the most massive supermassive black holes ($M_{\rm BH}$ > 10$^8$ $M_{\odot}$) in their centres \citep{2000laor1}. The question of how the jets are triggered was, and still remains, a subject of debate \citep{2016tadhunter1}. Some authors believe that galaxy interactions and mergers are connected to the triggering of relativistic jets, but the physical mechanism behind the triggering remains unclear. There are two main scenarios that are typically considered in such discussions. In the first scenario, the jet draws its energy from the spin of the black hole via a magnetised accretion disk \citep{1977blandford1}. A black hole can be efficiently spun up in galaxy mergers when the central black holes coalesce \citep[e.g. ][]{2015chiaberge1} or this can happen via normal accretion over a longer time period. In the second scenario, the jet gets its energy directly from the accretion disk. Gas infall into the nucleus caused by galaxy interactions can lead to super-Eddington accretion which, in turn, can trigger the jet \citep{2015sadowski1}. However, there are also several studies that have not found any connection between interaction and jets \citep{2000corbin1,2011cisternas1,2012kocevski1}.

Either way, we know that there is a prominent connection between the AGN and its host galaxy \citep{2000ferrarese1}. Especially with regard to galaxies with classical bulges, the black hole mass is directly connected to the stellar velocity dispersion of the bulge and exhibits signs of co-evolution \citep{2013kormendy1}. At the same time, the central black hole feeds on the gas reservoirs of its host galaxy and the nuclear activity shapes the galaxy via the so-called AGN feedback \citep[e.g. winds, outflows, and jets; see][]{2018husemann1}.

The conventional jet paradigm was challenged when proof of relativistic jets in a class of young AGN \citep{2000mathur1,2015jarvela1} called narrow-line Seyfert 1 galaxies \citep[NLS1, ][]{1985osterbrock1}, was found \citep{2009abdo2}. Indeed, these sources generally reside in late-type galaxies and host black holes of moderate masses ($M_{\rm BH}$ < 10$^8$ $M_{\odot}$). They have since then been established as the newest class of gamma-ray emitting AGN with relativistic jets \citep{2015foschini1, 2016berton1, 2017berton1}. NLS1s reignited the discussion about the role of interaction in triggering the relativistic jets since a clear majority of the jetted NLS1s studied so far are found in interacting systems \citep{2008anton1, 2018jarvela1, 2019berton1, 2020olguiniglesias1}. This suggests that, at least in these sources, interaction plays a major role.

While it is becoming evident that a connection between the triggering of the jets and interaction exists, we still do not understand the necessary conditions for an interaction to lead to the launch of relativistic jets. If mere interaction would be sufficient to achieve this result, we should be able to observe this phenomenon in a variety of interacting galaxies. Instead, so far, it seems that only blazars and NLS1s are capable of producing fully developed gamma-ray emitting relativistic jets. This could be partly due to the concentration of resources focussed on the study of the most outstanding sources and partly due to observational biases (e.g. occasionally flaring sources are not included in $Fermi$-LAT catalogues; see \citealp{2018lahteenmaki1}). In this paper, we report the discovery of the first non-local intermediate-type Seyfert galaxy with prominent gamma-ray emission produced by relativistic jets. Throughout the paper, we assume a cosmology with H$_{0}$ = 73 km s$^{-1}$ Mpc$^{-1}$, $\Omega_{\text{matter}}$ = 0.27, and $\Omega_{\text{vacuum}}$ = 0.73 \citep{2011komatsu1}.

\section{Reclassification of SDSS J211852.96-073227.5}
\label{sec:reclass}

SDSS J211852.96-073227.5 (henceforth, J2118-0732) is a Seyfert galaxy at a redshift of 0.2601 \citep{2017albareti1}. It has been previously classified as an NLS1 by several authors \citep[e.g. ][]{2017rakshit1,2018yang1,2018paliya1} and it was identified as a gamma-ray emitter in the $Fermi$-LAT 8-year Source Catalog \citep[4FGL, ][]{2019fermi1} as well as in \citet{2018paliya1}.

To confirm the spectral classification of J2118-0732, we observed it on 2019-10-17 with the Low-Dispersion Survey Spectrograph 3 (LDSS3) mounted on the 6.5m Clay Telescope of the Las Campanas Observatory. We used grism VPH-ALL, with a spectral range of 4000-10000\AA{} and spectral resolution of R$\sim$650. We obtained three exposures of 900s, for a total exposure of 2700s. The 1$^{\prime\prime}$ slit was oriented along the parallactic angle. We reduced the spectra following the standard IRAF\footnote{IRAF is distributed by the National Optical Astronomy Observatories, which is operated by the Association of Universities for Research in Astronomy, Inc. under cooperative agreement with the National Science Foundation.} procedure. The signal-to-noise ratio (S/N) of the final combined spectrum is $\sim$30 in the continuum. These new deep spectroscopic observations show that this source is not a type 1 AGN but, rather, it is an intermediate AGN. This class of often-neglected AGN is characterised by the partial obscuration of the central engine and by H~I line profiles in which both the broad and the narrow component can be easily recognised \citep{1977osterbrock1}. 

\begin{figure}
    \centering
    \includegraphics[width=\hsize]{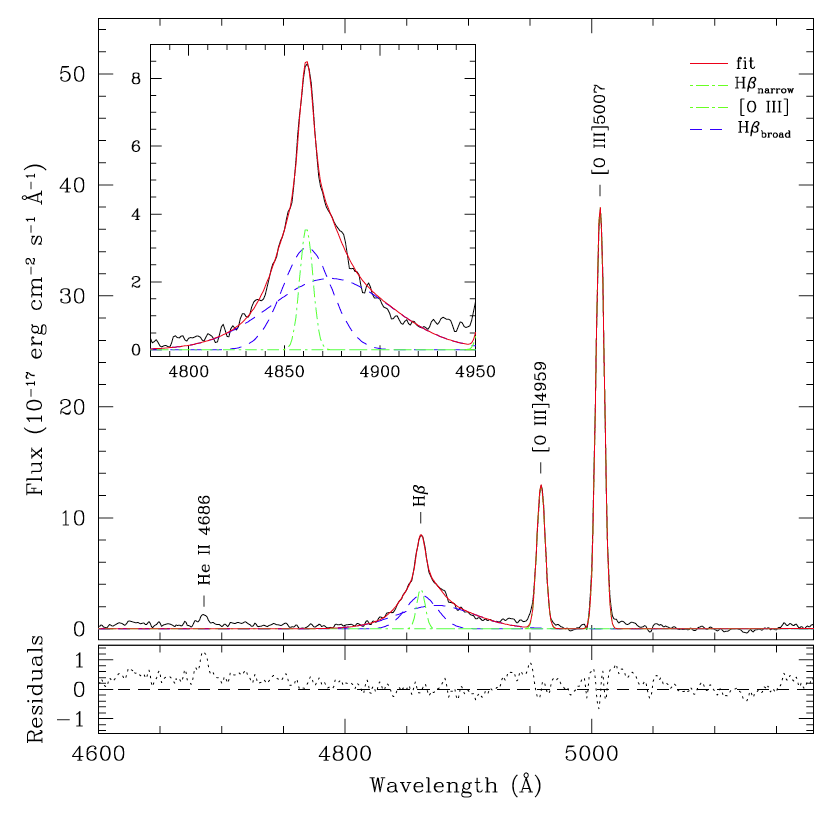}
    \caption{H$\beta$ and [O~III]$\lambda\lambda$4959,5007 of J2118-0732 at rest-frame, and a zoom-in to the H$\beta$ line. The black solid line represents the spectrum. Other symbols and colours are explained in the figure.}
    \label{fig:fit_hb}
\end{figure}

\begin{figure}
    \centering
    \includegraphics[width=\hsize]{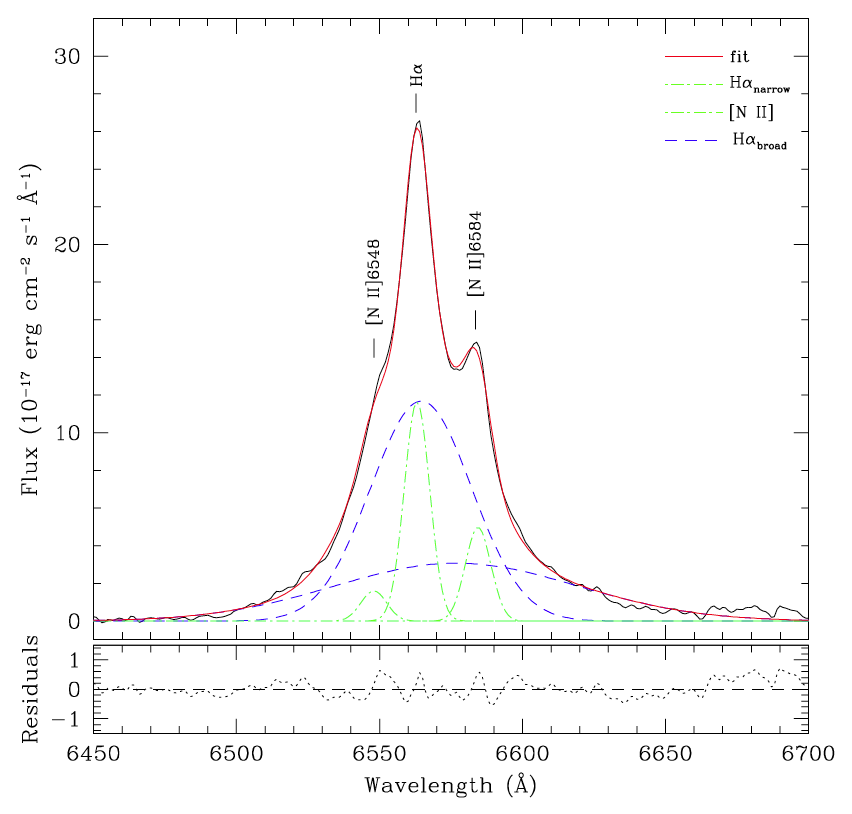}
    \caption{H$\alpha$ and [N~II]$\lambda\lambda$6548,6584 of J2118-0732 at rest-frame. Colours are the same as in Fig.~\ref{fig:fit_hb}}
    \label{fig:fit_ha}
\end{figure}

The strongest hydrogen lines of J2118-0732 are shown in Figs.~\ref{fig:fit_hb} and \ref{fig:fit_ha}. Both H~I line profiles are clearly asymmetric, showing a prominent wing on the red side. Nonetheless, we tried to reproduce the H$\beta$ profile by using a single Lorentzian component as is usually done for NLS1s \citep{1999sulentic1, 2016cracco1, 2020berton1}. Unlike what was found by \citet{2018yang1}, this function was not sufficient to reproduce the observed profile in this case. A Lorentzian component can reproduce only the wings of the profile or its core, but not both. We then added a narrow component to the fit, which significantly improved the result, but could not account for the red asymmetry. Therefore, we decided to use only Gaussians in the fitting procedure. We used two Gaussians to represent the broad component and one Gaussian for the narrow component. The result is shown in Fig.~\ref{fig:fit_hb}. The full-width at half maximum (FWHM) of the broad components is $\sim$4560 and $\sim$1860 km s$^{-1}$ , respectively, in the restframe (not corrected for instrumental resolution). In the narrow component, the FWHM is $\sim$490 km s$^{-1}$, which corresponds to the instrumental resolution and is in agreement with the values measured in the [O~III] lines. The same model was then successfully applied to the H$\alpha$+[N~II] blend, obtaining similar velocities as measured in H$\beta$ (see Fig.~\ref{fig:fit_ha}). In an NLS1, the line profile is usually dominated by the broad component(s), and the narrow component is almost negligible \citep{1999sulentic1}, whereas in the case of J2118-0732, the narrow component significantly contributes to the line profile and, thus, it cannot be ignored. For this reason, the NLS1 classification seems unlikely, while the intermediate-type classification is more apt for this source.

In their analysis, \citet{2018yang1} reported the presence of Fe~II multiplets, as often seen in NLS1s \citep[e.g.][]{1989goodrich1}. They estimated the Fe~II to H$\beta$ ratio -- which is known as R4570 -- to be 0.06, indicating that the iron emission is relatively weak. Our spectrum appears to confirm their findings since the Fe~II emission is barely visible. We did not try to fit the Fe~II multiplets; its contribution can be seen as residuals below 4800\AA \hspace*{0.05cm} in Fig.~\ref{fig:fit_hb}. However, this does not contradict the classification of J2118-0732 as an intermediate Seyfert since Fe~II multiplets are also observed in intermediate-type AGN \citep[e.g. NGC 5548,][]{2005vestergaard1}.

The reason for the discrepancy between our analysis and that of \citet{2018yang1} is due to the divergence in S/N. Their results were based on the Sloan Digital Sky Survey (SDSS) spectrum, with an S/N$\sim$10 in the continuum. Our spectrum, instead, has a significantly better S/N of $\sim$30, allowing us to study the line profiles in more detail. 

\section{Host galaxy}
\label{sec:host}

To study the morphology of J2118-0732, we observed it in $J$ and $Ks$ -bands using the FourStar instrument on the 6.5m Baade Telescope of the Las Campanas Observatory. The observations were performed on 2019-10-10 and the seeing varied between 0.5 and 0.8$^{\prime\prime}$. The spatial scale of the images is 0.16$^{\prime\prime} px^{-1}$, which, at the redshift of the target, translates to 4.048 kpc/$^{\prime\prime}$. In the $J$-band, we observed the target in 11 different positions on the detector, combining eight images for each position, with a 14.555s exposure time of a single frame and a total exposure of $\sim$1280s. In the $Ks$-band, we observed the target in 11 different positions with 12 exposures per position. The single exposure was 5.822s, and the total exposure time was $\sim$768s. We reduced the images with IRAF following the standard imaging procedure. We flux-calibrated the final images, using the magnitudes of the field stars derived from 2MASS as reference \citep{2003cutri1}. 

\begin{figure*}[ht!]
\centering
\adjustbox{valign=t}{\begin{minipage}{0.372\textwidth}
\centering
\includegraphics[width=1.0\textwidth]{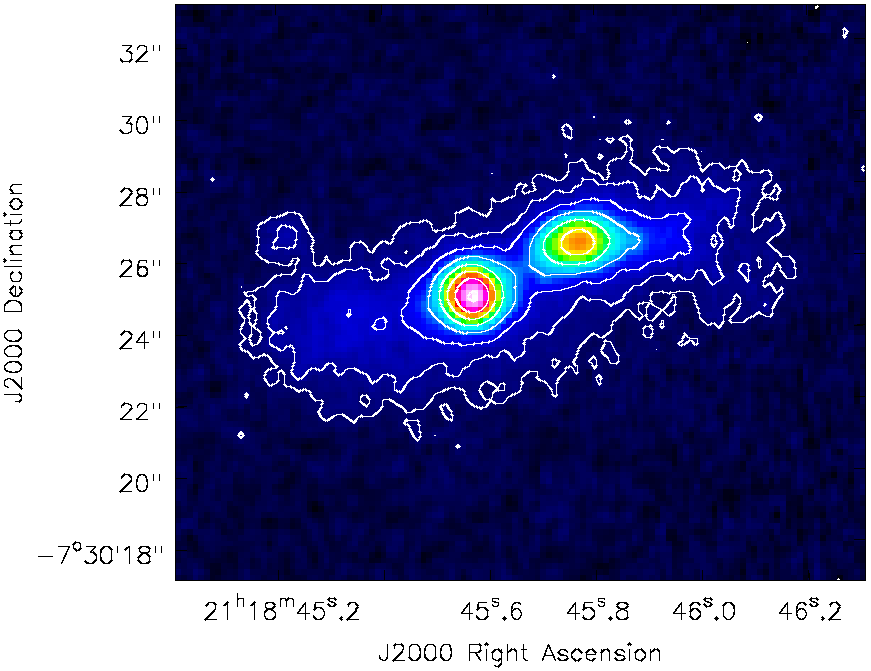}
\end{minipage}}
\adjustbox{valign=t}{\begin{minipage}{0.308\textwidth}
\centering
\includegraphics[width=0.98\textwidth]{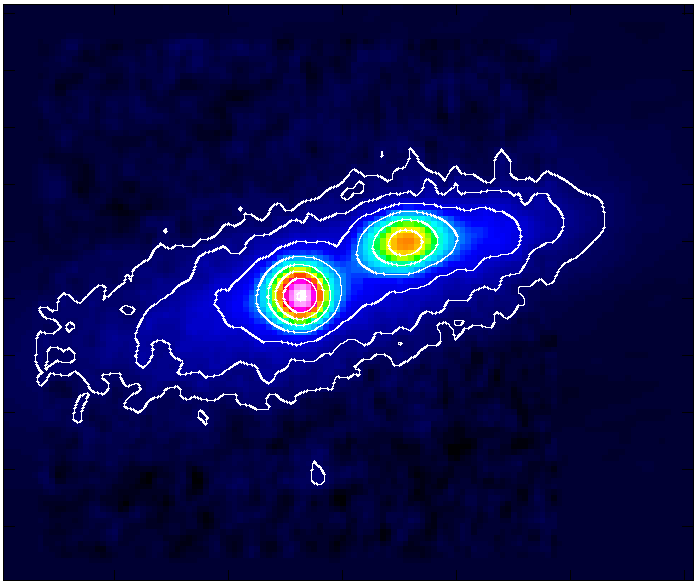}
\end{minipage}}
\adjustbox{valign=t}{\begin{minipage}{0.308\textwidth}
\centering
\includegraphics[width=0.98\textwidth]{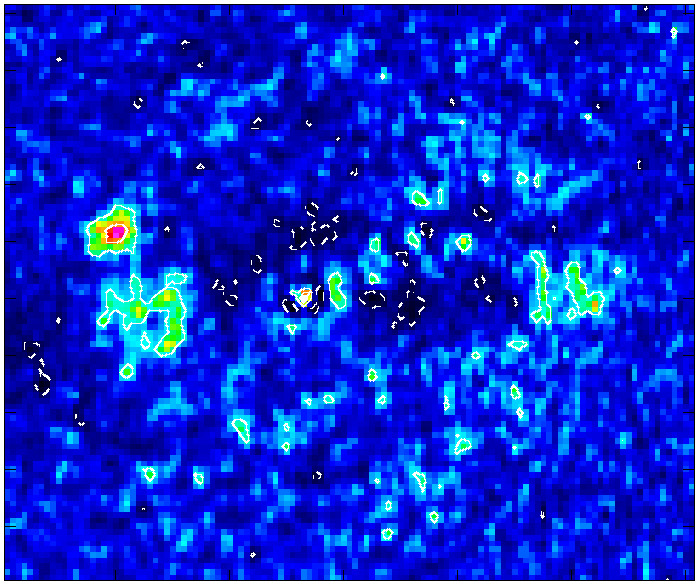}
\end{minipage}}
\hfill
\caption{Observed (left panel), modelled (middle panel), and residual (right panel) $Ks$-band image of J2118-0732. The contours are at -3, 3, 6, 12, 24, 48, 96, 192, and 384 $\times$ the noise level. The field of view is 19.36" / 78.4~kpc for all images. } \label{fig:2118}
\end{figure*}

\begin{figure} 
    \centering
    \includegraphics[width=\hsize]{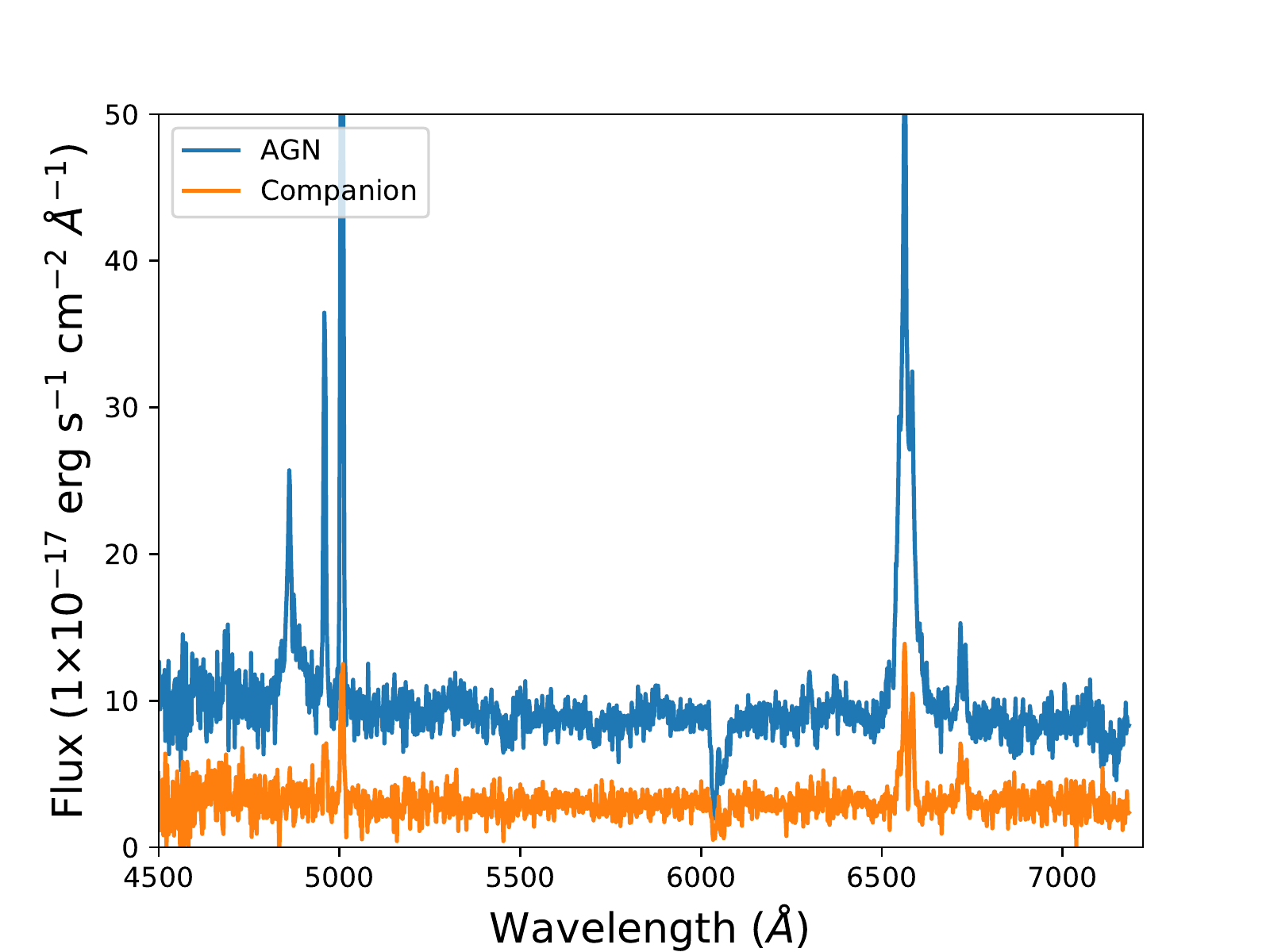}
    \caption{Redshift-corrected spectra of J2118-0732 (in blue) and of its companion (in orange).}
    \label{fig:redshift_companion}
\end{figure}

It is clear from the $Ks$-band image of J2118-0732 (see Fig.~\ref{fig:2118}, left panel) that it resides in an interacting system. To confirm this, we obtained the spectra of J2118-0732 and its companion with the Alhambra Faint Object Spectrograph and Camera (ALFOSC) instrument of the Nordic Optical Telescope (NOT) on 2019-09-04. We observed the system using the 1$^{\prime\prime}$ slit, grism\#20 (spectral coverage 5650-10150\AA, resolution R$\sim$770). We obtained three exposures of 900s each, for a total exposure time of 2700s. The slit was oriented at a position angle of 118$^\circ$ to observe both targets simultaneously. We reduced the spectra following the standard procedure with IRAF. The result, corrected for redshift, is shown in Fig.~\ref{fig:redshift_companion}. It confirms that the AGN and its companion are at the same redshift. The AGN resides in the brighter galaxy on the left in Fig.~\ref{fig:2118}.

\renewcommand{\arraystretch}{1.5}
\begin{table*}[ht!]
\caption[]{Best fit parameters of J2118-0732 and its companion in $Ks$-band. Reduced $\chi^2_{\nu}$ = 1.37 $\substack{+0.02 \\ - 0.02}$ .}
\centering
\begin{tabular}{l l l l l l l}
\hline\hline
funct.       &  mag                            & $r_{e}$                         & $n$                            & axial                          & PA                            & notes   \\
             &                                 & (kpc)                           &                                & ratio                          & (\textdegree)                 &  \\ \hline
$AGN$        &                                 &                                 &                                &                                &                               &   \\
PSF          & 20.09 $\substack{+0.18\\-0.08}$ &                                 &                                &                                &                               & nucleus\\
S\'{e}rsic 1 & 16.05 $\substack{+0.06\\-0.07}$ & 21.25 $\substack{+0.85\\-0.66}$ & 0.79 $\substack{+0.08\\-0.07}$ & 0.31 $\substack{+0.00\\-0.00}$ & -75.2 $\substack{+0.0\\-0.0}$ & disk \\
S\'{e}rsic 2 & 16.44 $\substack{+0.00\\-0.00}$ & 0.56 $\substack{+0.01\\-0.01}$  & 1.90 $\substack{+0.06\\-0.08}$ & 0.82 $\substack{+0.01\\-0.01}$ & -75.8 $\substack{+0.1\\-0.1}$ & bulge \\
$companion$  &                                 &                                 &                                &                                &                               &   \\
S\'{e}rsic 3 & 17.10 $\substack{+0.00\\-0.00}$ & 2.22 $\substack{+0.00\\-0.00}$  & 1.07 $\substack{+0.01\\-0.00}$ & 0.40 $\substack{+0.00\\-0.00}$ & -79.9 $\substack{+0.0\\-0.1}$ & bulge \\
S\'{e}rsic 4 & 18.27 $\substack{+0.01\\-0.00}$ & 7.22 $\substack{+0.02\\-0.05}$  & 0.66 $\substack{+0.00\\-0.01}$ & 0.35 $\substack{+0.00\\-0.00}$ & -75.2 $\substack{+0.0\\-0.0}$ & off-centre disk?\\ \hline
\end{tabular}
\tablefoot{(1) Function used in the model; (2) Magnitude of the component in $Ks$-band; (3) Effective radius; (4) S\'{e}rsic index; (5) Axial ratio; (6) Position angle; (7) Physical interpretation.}
\label{tab:j2118}
\end{table*}

We used the $Ks$-band image to model the system because the weather conditions were better during these observations and the $Ks$-band image is deeper than the $J$-band image. Details of the modelling are given in Appendix~\ref{app:imaging-main} and the best-fit parameters of the morphology of the system are listed in Table~\ref{tab:j2118}. In $Ks$-band the AGN nucleus is quite faint which is not surprising since generally the host galaxy starts to dominate more and more at longer wavelengths \citep{2017hernancaballero1,2015caccianiga1}. The overall morphology of the system is significantly disturbed by interaction, which makes modelling individual components of the two galaxies challenging as their light distributions are no longer intact. However, the galaxy hosting the AGN can best be modelled with two S\'{e}rsic functions, which are concentric with the point spread function (PSF). S\'{e}rsic 1, with effective radius, $r_e$ = 21.25~kpc, and S\'{e}rsic index, $n$ = 0.79, resembles a disk component. The axial ratio is quite small (0.31), possibly due in part to the inclination of the disk, but also because this component includes most of the diffuse light of the whole system, which extends to the companion. S\'{e}rsic 2, with $r_e$ = 0.56~kpc, $n$ = 1.90, and an axial ratio of 0.82, fits the description of a pseudo-bulge \citep[e.g. ][]{2008fisher1}, confirming that prior to the interaction, the AGN host galaxy was of a late-type.

The companion can also be fit with two S\'{e}rsic functions, however, based on visual inspection, it seems that its morphology is already more perturbed than that of the galaxy hosting the AGN. It is most probably due to this that the two S\'{e}rsic functions are not concentric: trying to freeze the central coordinates leads to non-physical parameter values, so we decided to keep them free. S\'{e}rsic 3, with $r_e$ = 2.22~kpc and $n$ = 1.07, resembles a pseudo-bulge. However, it is quite elongated, with axial ratio of 0.40, which is probably due to the interaction. The second companion component, S\'{e}rsic 4, is off-centre by 9.02~kpc when compared to the bulge, and clearly fits the elongated structure extending to the right side of the companion bulge. S\'{e}rsic 4 has $r_e$ = 7.22~kpc and $n$ = 0.66, which suggests that it is a disk-like structure. 

The projected separation of the two nuclei is 13.36~kpc, which suggests that the system might be in a pre-merger stage and not just experiencing interaction when passing each other \citep{2018silva1,2019ventou1}. However, we do not know the de-projected separation or the velocity difference of the galaxies and, thus, we cannot be certain of the future evolution of the system.

\begin{figure}
    \centering
    \includegraphics[width=9cm]{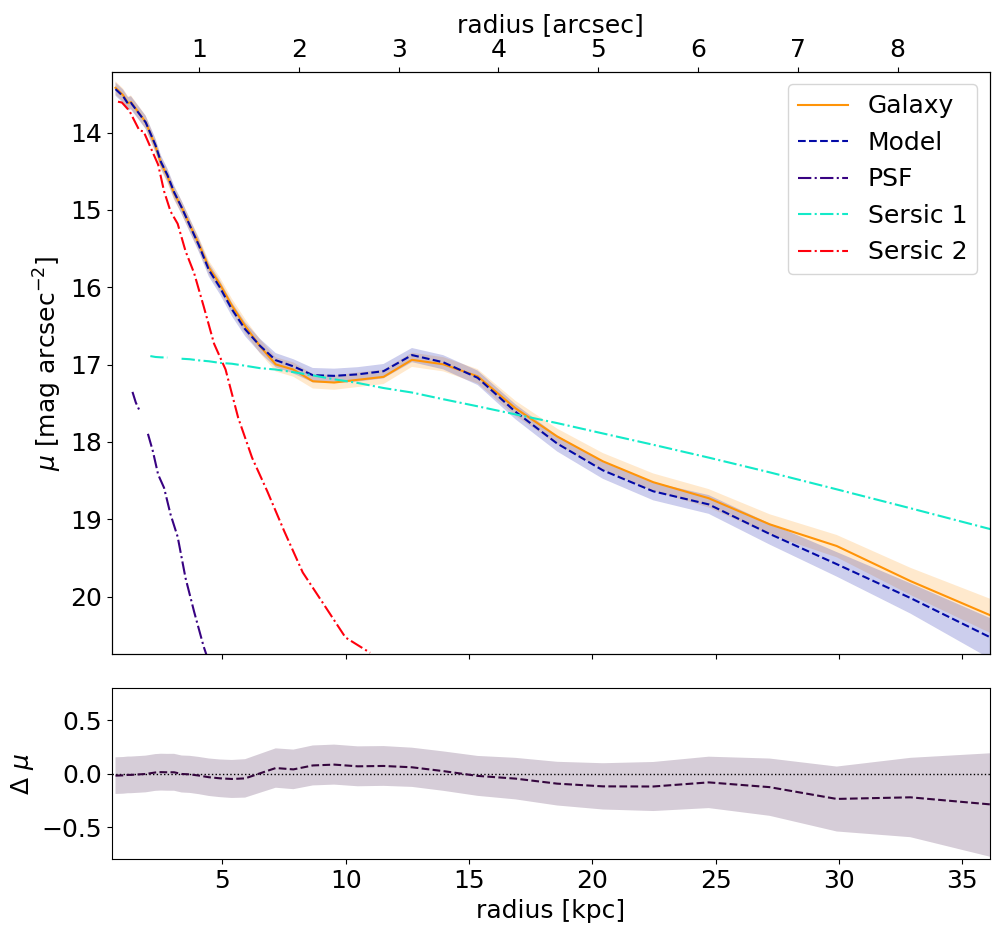}
    \caption{Radial surface brightness distribution of the system in $Ks$-band. Symbols and colours are explained in the plot and in the text. The shaded area describes the associated errors.}
    \label{fig:surf}
\end{figure}

The observed image, the model, and the residuals are shown in Fig.~\ref{fig:2118}. The radial surface brightness distribution (centred at the AGN) of the original image, the model, and the AGN components are shown in Fig.~\ref{fig:surf}. The companion galaxy is responsible for the bump at $r\sim$13~kpc, but we do not show its components since they are not concentric with the AGN. At large radii the disk component (S\'{e}rsic 1) seems, misleadingly, to have a higher surface brightness than the whole profile, but this is due to it being more elliptical than the total profile and due to the plotting the surface brightness as a function of the semi-major axis.

Our model fits the inner parts of the system well, but residuals are left at larger radii. These structures probably result from the interaction and we did not even try to fit them. Despite the somewhat disturbed morphology of the outer parts of the interacting galaxies, the best-fit indicates that both galaxies have pseudo-bulges and were originally late-type galaxies. 


\begin{figure}
    \centering
    \includegraphics[width=9cm]{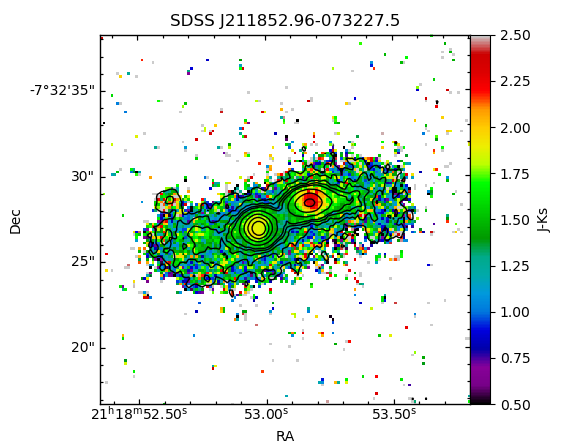}
    \caption{$J-Ks$ colour-colour image of the system.}
    \label{fig:colour}
\end{figure}

The $J-Ks$ colour-colour image of the system is shown in Fig.~\ref{fig:colour}. The colour of the companion nucleus is very red ($J-Ks$ $\sim$2.3), which may indicate strong absorption due to dust at that position. In the infrared, there may also be a contribution of stellar populations at play since old stars can contribute to the colours. However, using the Balmer decrement in the NOT spectrum, we estimated H$\alpha$/H$\beta \simeq$ 8, which corresponds to A(V) $\sim$3.2. Therefore, dust is believed to be the main suspect producing this feature \citep{2013dominguez1}.

\section{Discussion}
\label{sec:disc}

As we mention above, J2118-0732 is a gamma-ray source. Given its relatively high redshift and variable gamma-ray emission \citep{2018yang1}, the most likely source of the high-energy emission is a beamed relativistic jet \citep{2012ackermann1}. Gamma-ray emission in galaxies can also originate from star formation processes, but the emission produced this way is quite faint and, thus, it is difficult to detect in non-local galaxies; furthermore, it is not variable. The host galaxy of J2118-0732 is clearly interacting with a nearby companion galaxy. The morphology of the system is already quite disturbed, but based on our modelling, both galaxies have pseudo-bulges, indicating that both of them had late-type morphologies before the encounter. This result is in agreement with the postulated connection between interaction and the presence of relativistic jets. With these data, however, we can not assess the triggering mechanism of the jet since we do not know the accretion history of the AGN.

Beside the  $\sim$20 gamma-ray emitting NLS1s we know so far \citep[full list in][]{2018romano1}, relativistic jets, or gamma-ray emission have been very rarely observed in non-local Seyfert or late-type galaxies. For example, \citet{2012ackermann1} investigated the gamma-ray properties of 120 hard X-ray-selected Seyfert galaxies, but did not find any certain detections. Moreover, the two candidate gamma-sources are both nearby galaxies. The Seyfert 1.5 galaxy III Zw 2 has been found to host a gamma-ray emitting relativistic jet \citep{2000brunthaler1,2016liao1}, but according to the detailed modelling of the host galaxy it resides in an interacting elliptical \citep{2009veilleux1}. In addition, about a dozen late-type galaxies have been found to host kpc- or even Mpc-scale jets \citep{2011hota1,2014bagchi1,2015singh1} but their physical properties (e.g. black hole and stellar masses) are similar to those of blazars, BL Lac objects in particular, and the AGN hosted in these late-type galaxies are believed to be accreting very inefficiently \citep{2014bagchi1}.

In addition to modelling its host galaxy, we re-classified J2118-0732 as an intermediate-type AGN. This rather unexpected discovery properly demonstrates the dangers of not using high S/N, good quality data, as intermediate AGN can be easily misclassified. It also raises a question about the robustness of the optical classification of AGN in general since many of them have been classified based on quite low S/N SDSS spectrum and often using automated algorithms. Correct classification of AGN is crucial for population-wide studies since misclassified sources contaminate the samples, add noise to the results, and they can even lead to misleading results.

In general gamma-ray emission from AGN classified as intermediate is rare: in some local intermediate type AGN, such as NGC 4151 or MGC 6-11-11 \citep{1986ulvestad1}, gamma-ray emission was detected already with EGRET and later confirmed with $Fermi$-LAT, but its origin may reside in annihilation radiation from hot plasma instead of a relativistic jet \citep{1983bassani1,2015wojaczynski1}. High-energy emission from misaligned, that is, intermediate and type 2, AGN is not common, with only a handful of sources (mostly radio galaxies) detected by the \textit{Fermi}-LAT \citep{2010abdo2,2018rieger1}. However, in the case of J2118-0732, the relativistic jet does not seem to be misaligned with respect to our line of sight since it has been identified as a flat-spectrum radio source \citep{2007healey1}.

J2118-0732 is exceptional in many ways as it is the first confirmed case of a non-local intermediate Seyfert nucleus, hosted by an interacting late-type galaxy and capable of maintaining powerful relativistic jets that are detectable at gamma-ray energies. Its Eddington ratio, bolometric luminosity, and black hole mass have been estimated by previous authors but following the reclassification, we can not determine how accurate these estimates are. The intermediate classification means that the emission from the nucleus is partially obscured but we do not know by how much. In addition, the previous authors used FWHM(H$\beta$) (for which they got values < 2000 km s$^{-1}$) to estimate the black hole mass. This is clearly different from the values we obtained; either way, estimating the black hole mass from the line width of an intermediate AGN is not straightforward. Nonetheless, based on its prominent emission lines, we can deduce that it accretes via cold gas accretion and is in a high-excitation mode \citep{2006hardcastle1}, making it comparable to NLS1s and flat-spectrum radio quasars (FSRQ). It has been suggested that NLS1s will evolve into FSRQs with time and J2118-0732 could represent the transitional evolutionary phase \citep{2017berton1}.

Obscuration of the nucleus is an interesting feature of J2118-0732 and there are two possible explanations for it: obscuration by the dusty torus or by interstellar matter. In the simplest model an AGN is an axisymmetric structure, in which the dusty torus is the natural extension of the accretion disk beyond the dust sublimation radius and the jets are perpendicular to this system \citep{1993antonucci1, 1995urry1, 2009gaskell1}. However, various works have shown that the jets can get misaligned with respect to the obscuring material \citep{2013reynolds1,2013greene1}. This is believed to take place when the angular momentum vector of the accreted matter is not aligned with the axis of the obscuring medium. Such discrepancy may be due, for example, to galaxy interactions, which can disturb the gas dynamics of the participating galaxies. In general, in AGN, the half-opening angle of the torus is between 15\textdegree\hspace{0.02cm} and 55\textdegree, with a mean of 30\textdegree\hspace{0.02cm} \citep{1997petersonbook}. This means that the inclination of the jet relative to the dusty torus should be at least 15\textdegree\hspace{0.02cm} for us to observe the jet pointing toward us but also for the broad-line region to be partially obscured.

The other possibility is that the light emitted from the broad-line region is partially obscured by interstellar matter, especially dust, in the galaxy. It has been found that the central structure is often misaligned with the galaxy disk \citep[e.g.][]{2002schmitt1,2012hopkins1} and this can be responsible for the partial obscuring of the broad-line region. An alternative scenario is that the interaction has perturbed the distribution of the interstellar matter in J2118-0732, so that there is now obscuring dust in our line of sight to the nucleus. The current data are not sufficient to determine the source of the obscuration.

For now J2118-0732 remains a curiosity, however, with more careful data analysis in the advent of more sensitive instruments and higher quality data, it should not remain a mystery for much longer. Identifying more sources with properties that are similar to J2118-0732 is important in understanding their place in the big picture of AGN unification and evolution schemes. To achieve this and to investigate the nature of these sources, new observations, such as a combination of high-resolution radio imaging and integral field spectroscopy, will play a crucial role.

\begin{acknowledgements}
M.B and S.C. acknowledge the financial support from the visitor and mobility program of the Finnish Centre for Astronomy with ESO (FINCA), funded by the Academy of Finland grant nr 306531. The authors are grateful to Dr. Tuomas Savolainen for helpful discussion. This paper includes data gathered with the 6.5 meter Magellan Telescopes located at Las Campanas Observatory, Chile. Based on observations made with the Nordic Optical Telescope, operated by the Nordic Optical Telescope Scientific Association at the Observatorio del Roque de los Muchachos, La Palma, Spain, of the Instituto de Astrofisica de Canarias. This publication makes use of infrastracture at the Mets\"{a}hovi Radio Observatory, operated by Aalto University, Finland. The SDSS Web site is http://www.sdss.org/. The SDSS is managed by the Astrophysical Research Consortium (ARC) for the Participating Institutions. The Participating Institutions are The University of Chicago, Fermilab, the Institute for Advanced Study, the Japan Participation Group, The Johns Hopkins University, the Korean Scientist Group, Los Alamos National Laboratory, the Max-Planck-Institute for Astronomy (MPIA), the Max-Planck-Institute for Astrophysics (MPA), New Mexico State University, University of Pittsburgh, University of Portsmouth, Princeton University, the United States Naval Observatory, and the University of Washington. This research has made use of the NASA/IPAC Extragalactic Database (NED) which is operated by the Jet Propulsion Laboratory, California Institute of Technology, under contract with the National Aeronautics and Space Administration.
\end{acknowledgements}

\bibliographystyle{aa}
\bibliography{artikkeli.bib}

\appendix

\section{Image modelling}
\label{app:imaging-main}

To perform the photometric decomposition of the $Ks$-band image, we used GALFIT version 3 \citep{2010peng1}, which allows simultaneous fitting of several components that contribute to the total 2D light distribution of the source. To properly model the AGN, and remove its contamination, a good PSF model is needed. For this purpose, we used a nearby field star -- with as similar a PSF as possible -- that is more than two magnitudes brighter than the AGN and with an S/N of several hundred. GALFIT is capable of extracting the PSF directly from the PSF star image once it is centred and sky-subtracted. This allows more accurate PSF modelling since no analytical functions are needed to model the PSF.

We estimated the zeropoint of the image and its associated error by using stars in the field of view with 2MASS magnitudes in $Ks$-band. The image was sky-subtracted to have a mean of zero. The sky error was estimated by measuring the sky in several separate regions of 100$\times$100~px. In the fitting we froze the sky to have a value of zero and estimated the sky error effect by fitting with the $\pm$1$\sigma$ values. For magnitudes an additional zeropoint estimation error was added. The modelling was started by fitting only the PSF and then adding more components one at a time as required. The residuals were checked after every fit, and the component parameters were varied to ensure the values that the fit converged to were stable. We fixed the central coordinates of the AGN components but kept other parameters free. After achieving a good fit, determined by the reduced $\chi^2_{\nu}$ and the visual inspection of the model and the residuals, we also visually checked all the subcomponents to confirm they looked physically reasonable. The fitting region used was as large as possible but a zoom-in in shown in Fig.~\ref{fig:2118} for easier inspection. 

The S\'{e}rsic profile was used to model the various components of our sources:

\begin{equation}
    I(r) = I_e \texttt{exp} \Bigg[ -\kappa_n \Bigg( \bigg( \frac{r}{r_e} \bigg)^{1/n} -1 \Bigg) \Bigg]
,\end{equation}

where $I(r)$ is the surface brightness at radius $r$, $\kappa_n$ is a parameter connected to the S\'{e}rsic index, $n$, so that $I_e$ is the surface brightness at the half-light radius, $r_e$ \citep{2005graham1}. By changing the S\'{e}rsic index, $n$, the S\'{e}rsic profile can be used to model varying light distributions in galaxies, for example, classical and pseudo-bulges, and early- and late-type morphologies. Smaller values of $n$ ($\lesssim$ 2) are associated with galaxies with late-type morphology and pseudo-bulges, and larger values of $n$ ($\gtrsim$ 4) with elliptical galaxies and classical bulges \citep{2005graham1}. 

After a good fit was achieved we extracted the radial surface brightness profile from the observed image, the model image, and the separate component images using IRAF task $ELLIPSE$, which fits concentric elliptical isophotes to a 2D image. For the original image and the model image we used similar $ELLIPSE$ parameters to get comparable fits. For the individual components, we took the values for the central coordinates, the axial ratio, and the position angle from the GALFIT best-fit parameters. The error estimation takes into account the most important error sources, the sky value error, and the zeropoint error.
\end{document}